\documentclass[aps,twocolumn]{revtex4}
\usepackage{epsf}
 
\begin{document}
      \title{Correlation effects in the transport through quantum dots}

     \author{A.\ Donabidowicz, T.\ Doma\'nski, K.I.\ Wysoki\'nski}
     
\affiliation{
             Institute of Physics and Nanotechnology Centre,\\ 
	     M.\ Curie Sk\l odowska University,
             20-031 Lublin, Poland} 
      \date{\today}

\begin{abstract} 
We study the charge and heat transport through the correlated quantum 
dot with a finite value of the charging energy $U\!\neq\!\infty$ . The
Kondo resonance appearing at temperatures below $T_{K}$ is responsible 
for several qualitative changes of the electric and thermal transport. 
We show that under such conditions the semiclassical Mott relation between 
the thermopower and electric conductivity is violated. We also analyze 
the other transport properties where a finite charging energy $U$ has 
a  significant influence. They are considered here both, in the limit 
of small and for arbitrarily large values of the external voltage 
$eV\!=\!\mu_L\!-\!\mu_R$ and/or temperature difference $T_L\!-\!T_R$. 
In particular, we check validity of the Wiedemann-Franz law and the
semiclassical Mott relation.  
\end{abstract}

\maketitle

Recent development in nanoelectronics has renewed the interests 
for studying the correlation effects in systems where the localized 
quantum states coexist and interact with a sea of itinerant 
electrons. Previously some related ideas have been explored 
in the solid state physics when investigating the screening 
of magnetic impurities hybridized with the conduction band 
electrons \cite{Kondo-64}. Correlations lead there to the Kondo 
effect \cite{Hewson} which shows up by a logarithmic increase 
of the electrical resistance at temperatures below $T_{K}$. 
Similar mechanism has been later found to be responsible for a
superconductivity of the heavy fermion compounds \cite{heavy_f}. 

In the nano-  and mesoscopic systems (such as quantum dots or 
carbon nanotubes connected to external leads) there can also arise 
the Kondo effect \cite{Phys_world}. This has been predicted 
theoretically \cite{Lee,Glazman} and observed experimentally 
in measurements of the differential conductance $G(V)\!=\!
\frac{dI(V)}{dV}$ \cite{Goldhaber-98}. Recent experimental 
studies of the thermoelectric power $S \!=\! - \frac{V}{\Delta T}
\;_{| I(V)=0}$  \cite{Scheibner-05} provided an additional 
evidence in favor of the Kondo effect at low temperatures .

Various transport properties through the correlated quantum dots 
have been studied in quite a detail \cite{Glazman,Lee,
Goldhaber-98,Franceschi,Scheibner-05,Simmel,Schmid,Kim,Meir-92}. 
Authors investigated the quantum dots coupled to the normal, 
superconducting and ferromagnetic leads. In several cases there 
have been observed some qualitatively new effects, for instance 
a splitting of the zero bias resonance under magnetic field 
\cite{Meir-92,Goldhaber-98}, absence of the even-odd parity 
effects \cite{Schmid} or the out of equilibrium Kondo effect 
\cite{Simmel,Krawiec-02,Franceschi}. Kondo effect has been 
found in such nanosystems like the single atom \cite{Park}, 
single molecule \cite{Liang} and carbon nanotubes \cite{Nygard}. 
Moreover, it has been also observed in the quantum dots attached 
to ferromagnetic leads \cite{Pasupathy} where, in principle, 
transport is controlled by the spin degree of freedom. 

Quantum dots are often thought to be promising objects for 
the nanotechnology e.g.\ in various spintronics applications 
\cite{Prinz}, as the gates of quantum computers, 
in electron spin entanglers \cite{Fabian} or efficient energy 
conversion devices \cite{Heremans}. Therefore a detailed 
understanding of the charge and heat transport through 
mesoscopic systems seems to be of primary importance \cite{TEP}. 
Besides the differential conductance it is important to analyze 
on equal footing also the thermoelectric properties because they 
are very sensitive to dimensionality as well as the electronic 
spectrum near the Fermi level. In the Coulomb blockade regime 
the thermoelectric coefficient $S$ has been shown \cite{Andreev01} 
to oscillate as a function of the gate voltage (a characteristic 
saw-tooth shape). At temperatures of the order or smaller than 
$T_K$ there forms a narrow resonance slightly above the Fermi 
energy. Its appearance is responsible for increasing the differential 
conductance (the zero bias anomaly) and simultaneously leads to 
a change of sign of the thermopower from positive (when 
$T>T_{K}$) to negative values (for $T$ smaller than $T_{K}$). 
Both these effects are useful for detecting the Kondo effect 
in the quantum dots \cite{Boese-01,Kim}. 

In most of the experimental measurements it has been found that 
the charging energy (the on-dot Coulomb repulsion) $U$ is large, yet 
being finite. Usually it varies between $2$ and $5$ in units of 
the effective coupling $\Gamma$ whose meaning will be explained 
in the Section II.  

It is a purpose of our work here to focus on studying the transport 
properties for the quantum dot with a finite Coulomb energy $U$. 
We analyze the linear response with respect to the bias $V$ and 
temperature difference $\Delta T$ applied across the junction 
and discuss some nonlinear effects in the case of arbitrary $V$ 
and $\Delta T$ values. Transport coefficients are determined with
use of the non-equilibrium Keldysh Green's functions 
\cite{Keldysh_formalism}. We treat the correlations 
using the equation of motion approach \cite{Caroli-71} whose 
justification has been recently thoroughly discussed in the 
Ref.\ \cite{Kashcheyeva}.

Studying the equilibrium and the non-equilibrium aspects of the 
electric and thermal transport we shall check a validity of the 
Wiedemann-Franz law and the Mott relation. The paper is organized 
as follows. In Section II we briefly introduce the model and the 
approximation used for calculating the transport coefficients. 
Results obtained within the linear response theory are presented 
in Section III. In the next Section IV we discuss a non-equilibrium 
case corresponding to arbitrarily large values of $V$ and $\Delta T$. 
We enclose this paper by a summary and some overview of the future 
prospects.

\section{Formulation of the problem}

\subsection{The model}

To account for the correlation effects we focus on the simplest
situation where QD is coupled to the normal leads as described 
by the non-degenerate single impurity Anderson Hamiltonian
\begin{eqnarray}
H & = & 
\sum_{{\bf k},\beta,\sigma }\xi _{{\bf k}\beta}
c^{+}_{{\bf k} \beta \sigma} c_{{\bf k} \beta \sigma}+
\sum_{\sigma} \epsilon_{d} d^{+}_{\sigma}d_{\sigma} 
+  U n_{d\uparrow} n_{d\downarrow} \nonumber \\
& + & \sum_{{\bf k},\beta, \sigma} \left( 
V_{{\bf k} \beta} c_{{\bf k} \beta \sigma}d_{\sigma}^{+}
+ V_{{\bf k} \beta}^{*} c^{+}_{{\bf k} \beta \sigma}
d_{\sigma} \right). 
\label{model}
\end{eqnarray}
Operators $c_{{\bf k} \beta \sigma}$ ($c_{{\bf k} \beta \sigma}^{+}$) 
annihilate (create) electrons in the left ($\beta\!=\!L$) or the 
right h.s.\ ($\beta\!=\!R$) electrodes with the corresponding 
energies $\xi_{{\bf k} \beta \sigma}\!=\!\epsilon_{{\bf k} \sigma}
\!-\!\mu_{\beta}$ measured from the chemical potentials shifted 
by the external voltage $\mu_{L}\!-\!\mu_{R}\!=\!eV$. Operators 
$d_{\sigma}$, $d_{\sigma}^{+}$ refer to the localized electrons 
of the dot which is characterized by a single energy level 
$\epsilon_{d}$ as well as by the Coulomb potential $U$. The 
last term in (\ref{model}) describes hybridization between 
the localized and itinerant electrons.

The retarded Green's function of the QD can be expressed 
in a general form
\begin{eqnarray}
G_{d}^{r}(\omega) = \frac{1}{\omega - \varepsilon_{d}
-\Sigma_{0}(\omega) - \Sigma_{I}(\omega)}.
\label{retarded_GF}
\end{eqnarray}
It consists of two contributions $\Sigma_{\nu}(\omega)$ to the 
selfenergy, where the first part $\Sigma_{0}(\omega)=\sum_{{\bf k},
\beta}\frac{|V_{{\bf k}\beta}|^{2}}{\omega - \xi_{{\bf k}\beta}}$ 
corresponds to the noninteracting case $U\!=\!0$. Twice of 
its imaginary part is often used as a convenient definition 
of the effective coupling $\Gamma_{\beta}(\omega)\!=\!2\pi 
\sum_{\bf k} |V_{{\bf k}\beta}|^{2} \delta(\omega\!-\!\xi_{
{\bf k}\beta})$. We assume a flat function $\Gamma_{\beta}
(\omega)=\Gamma_{\beta}$ for $|\omega|\leq D$ so, that the 
uncorrelated QD is characterized by the Lorentzian density of 
states centered around $\varepsilon_{d}$ with the halfwidth 
$\Gamma_{L}+\Gamma_{R}$.

\begin{figure}
\parbox{9cm}{
{\epsfxsize=4.3cm\epsffile{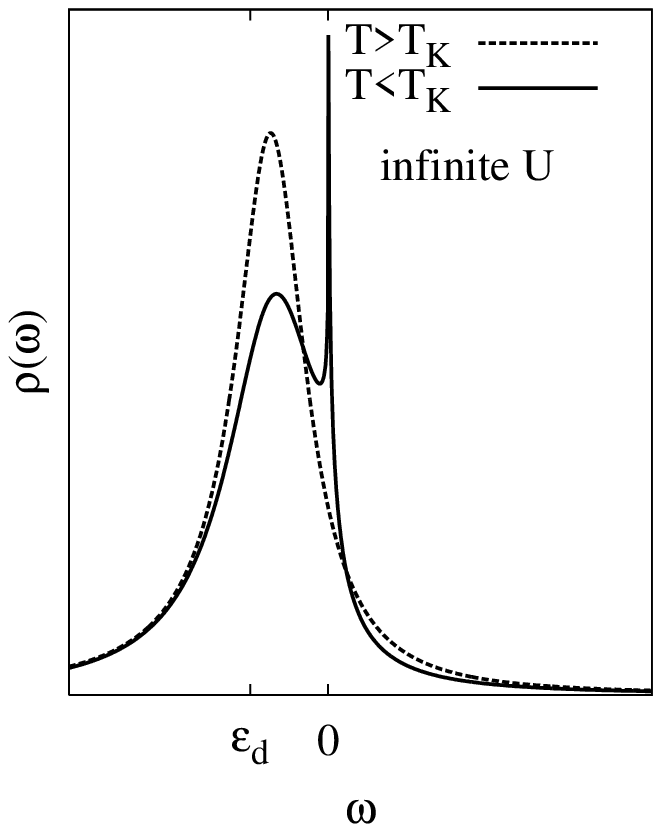}}\hspace{-0.3cm}
{\epsfxsize=4.3cm\epsffile{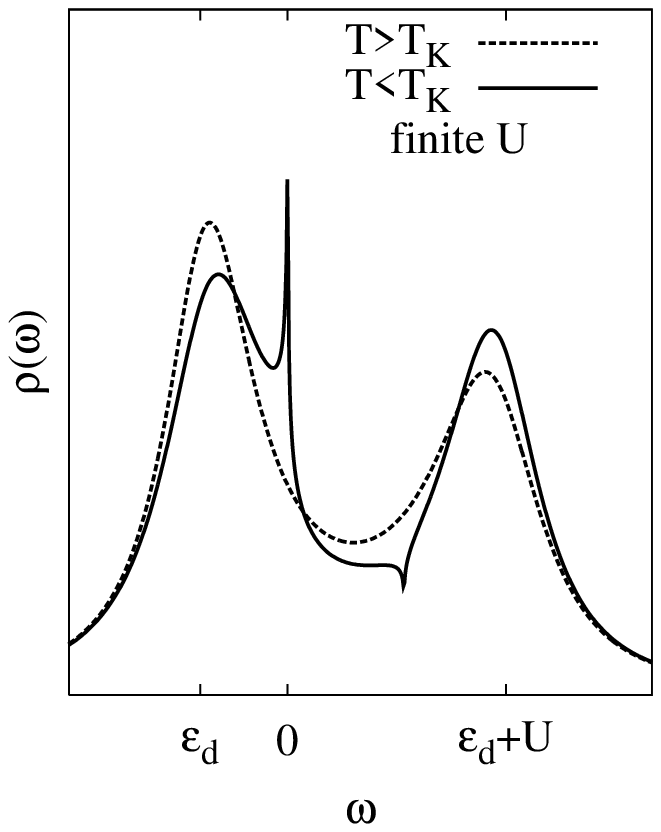}}}
\caption{Density of states $\rho_{d}(\omega)$ of the QD 
in the equilibrium case. For a finite Coulomb energy $U$ 
the spectrum consists of two Lorentians centered around 
$\epsilon_{d}$ and $\epsilon_{d}+U$. At sufficiently low 
temperatures $T\!<\!T_{K}$ there appears a narrow Kondo 
resonance near the Fermi level (here it corresponds to 
$\omega\!=\!0$).}
\label{intro}
\end{figure}

The real many-body problem is encountered in (\ref{retarded_GF}) 
via additional part $\Sigma_{I}(\omega)$ whenever $U\!\neq\!0$. 
In the extreme limit $U \rightarrow \infty$ one can use the 
slave boson approach which approximately yields the following 
Green's function \cite{infiniteU_EOM} $G_{d}^{r}
(\omega)=\left[ 1 - \langle n_{d} \rangle \right] / \left[ 
\omega \!-\! \varepsilon_{d} \!-\! \sum_{{\bf k},\beta}|V_{{\bf 
k}\beta}|^{2}\frac{1+f_{\beta}(\omega)}{\omega - \xi_{{\bf k}
\beta}} \right]$ where $f_{\beta}(\omega)=1/\left[ 1 + \mbox{exp}
(\xi_{{\bf k}_{\beta}}/k_{B}T) \right]$. For sufficiently 
low temperatures the local density of states $\rho_{d}( \omega ) 
= - \frac{1}{\pi} \mbox{Im} G_{d}^r(\omega+i0^{+})$ develops
a narrow Kondo peak at the Fermi level as illustrated in 
figure \ref{intro}.

%

One of the methods for studying the effects of Coulomb 
interactions $U$ uses a truncation scheme for a set of 
the coupled Green's functions \cite{finiteU_EOM}. This, so called,  
equation of motion (EOM) technique is approximate. In particular, 
it is known that such procedure underestimates both the width 
and height of the Kondo resonance \cite{infiniteU_EOM}. There 
are also other techniques like NCA or the large N expansion
but all of them have only a limited range of applicability 
depending on the values of $U$, $V_{{\bf k}\beta}$ as well 
as temperature $T$. In this work we simply use the EOM to 
explore a qualitative behavior of the transport properties 
for the system ({\ref{model}). Our results can be eventually 
quantitatively improved by the more sophisticated methods.
 
According to the standard derivation based on the EOM and 
the Keldysh formalism \cite{Haug-96} one obtains the following 
retarded Green's function (\ref{retarded_GF}) of the quantum 
dot \cite{finiteU_EOM}

\begin{widetext}
\begin{eqnarray}
G^{r}_{d}(\omega)& = & \frac{
g^{r}_{d}(\omega)^{-1} - [\Sigma_{0}(\omega)+
\Sigma_{3}(\omega)+U(1-n_{d})]}
{[g^{r}_{d}(\omega)^{-1}-\Sigma_{0}(\omega)]
[g^{r}_{d}(\omega)^{-1}-(U+\Sigma_{0}(\omega)
+\Sigma_{3}(\omega))]+U \Sigma_{1}(\omega)]} ,
\label{finiteU_GF} 
\end{eqnarray}
\end{widetext}
where  
\begin{eqnarray}
g^{r}_{d}(\omega) & = & \left[\omega-\epsilon_{d} 
\right]^{-1} \nonumber \\
\Sigma_{1}(\omega) & = &  \sum_{{\bf k} \beta} 
|V_{{\bf k} \beta}|^{2} \left( \frac{f_{\beta}(\omega)}
{\omega\! - \! \xi_{{\bf k} \beta}} + 
\frac{f_{\beta}(\omega)}{\omega\! -\! U \! - 2 
\varepsilon_{d}\!+\!\xi_{{\bf k} \beta}} \right) ,
\nonumber \\
\Sigma_{3}(\omega) & = &   \sum_{{\bf k} \beta} 
|V_{{\bf k} \beta}|^{2} \left( \frac{1}
{\omega \!-\!\xi_{{\bf k} \beta}} + \frac{1}
{\omega\! -\! U \!- 2 \varepsilon_{d}
\!+\! \xi_{{\bf k} \beta}} \right) .
\nonumber
\end{eqnarray}
In figure (\ref{intro}) we show the density of states
$\rho_{d}(\omega)$ computed for the equilibrium case ($V\!=\!0$ 
and $T_{L}=T_{R}$). The right h.s.\ panel illustrates the spectrum 
for the case of a finite $U$ which is characterized by two Lorentians, 
one at $\varepsilon_{d}$ and another one around $\varepsilon_{d}\!+\!U$. 
The upper Coulomb satellite is missing (see the left h.s.\ panel) 
when $U\!=\!\infty$. At sufficiently low temperatures $T<T_{K}$ 
there appears a narrow Kondo resonance due to the singlet state 
formed from the itinerant electrons and the localized electrons 
of the quantum dot.

\subsection{Definition of the transport coefficients}

By applying some bias $V$ or temperature imbalance $T_{L}
\! \neq \! T_{R}$ the system is driven out of its equilibrium 
\cite{Caroli-71}. Effectively there are induced the charge and 
heat currents through the QD. Using the nonequlibrium Keldysh 
Green's functions \cite{Keldysh_formalism} one can derive the 
following Landuer-type expressions for the charge $I(V,\Delta 
T)$ and heat currents $I_{Q}(V,\Delta T)$ \cite{Meir-92} 
\begin{eqnarray}
I &=&  \frac{e}{h} \int_{-\infty }^{\infty } d\omega 
\Gamma(\omega) \; [f_L(\omega)\!-\!f_R(\omega) ]
\; \rho_{d}(\omega), \label{charge_current} \\
I_{Q}&=& \frac{1}{h} \int_{-\infty }^{\infty } 
d\omega \Gamma(\omega) \; (\omega\!-\!eV) \; 
[f_L(\omega)\!-\!f_R(\omega) ]\; \rho_{d}(\omega) . 
\nonumber \\ & & \label{heat_current} 
\end{eqnarray}
We introduced here the symmetrized coupling $\Gamma(\omega)=
2\Gamma_{L}(\omega) \Gamma_{R}(\omega)/[\Gamma_{L}(\omega)
+\Gamma_{R}(\omega)]$ and factor $2$ comes from the 
contributions of $\uparrow$ and $\downarrow$ electrons 
in the tunneling. We notice that both currents 
(\ref{charge_current},\ref{heat_current}) are convoluted 
with the spectral function of the QD hence the behavior 
illustrated in figure \ref{intro} indirectly affects 
the transport properties.
 
In figure \ref{transport_fig2} we show the differential 
conductance $G(V)\!=\!dI/dV$ obtained for temperatures 
higher and lower than $T_{K}$. We notice a clear increase 
of the zero voltage conductance (zero bias anomaly) when  
$T<T_{K}$ while the other side-peaks in the conductance 
at $|eV| \simeq U$ are not sensitive to temperature. In 
realistic systems there should be even more peaks because 
of the multilevel electronic structure containing a 
mesoscopic number of atoms at QD. We thus emphasize 
that from these features only the zero bias anomaly is 
driven by the many-body Kondo effect at temperatures 
$T<T_K$, usually being of the order of hundreds mK
\cite{Goldhaber-98}.

For a further analysis we introduce the following transport 
coefficients   
\begin{eqnarray}
S & = & - \left( \frac{V}{\Delta T} \right)_{{I=0}} 
\label{S} \\
\kappa & = & - \left( \frac{I_{Q}}{\Delta T} \right)_{{I=0}} 
\label{kappa} 
\end{eqnarray}
which describe correspondingly the thermoelectric power 
(\ref{S}) and the heat conductance (\ref{kappa}). In the 
limit of small perturbations $V$ and $\Delta T$ the heat 
conductance is expected to obey the Wiedemann-Franz 
law $\kappa/T G(0)\!=\!\pi^{2}/3e^{2}$, however we will 
show that this relation is, in general, not obeyed for 
temperatures $T \sim T_{K}$ (as has been independently 
pointed out by several authors, e.g.\ \cite{Boese-01}). 

The thermoelectric power (\ref{S}) is a quantity which is 
very sensitive to a particle-hole symmetry in the density 
of states of QD. For small perturbations $V$, $\Delta T$ 
one can analytically prove that the thermopower is 
proportional to the energy weighted with respect to 
the Fermi level $\mu$
\begin{equation}
 S \simeq - \; \frac{\langle \omega-\mu \rangle}{eT} 
\label{SimpleFormFor_S}
\end{equation}
in the range of thermal excitations $k_{B}T$. Appearance 
of the narrow Kondo peak should thus be accompanied by 
a change of sign in $S(T)$ from the positive ($T>T_K$) 
to negative ($T<T_{K}$) values. Indeed, this behavior has 
been recently observed experimentally \cite{Scheibner-05}. 
In the following sections we discuss in a more detail 
some properties of the thermopower for a representative 
set of $U$ and for varying perturbations $V$, $\Delta T$ 
ranging from the small to the arbitrarily large values.

\begin{figure}
\centerline{\epsfxsize=8cm \epsffile{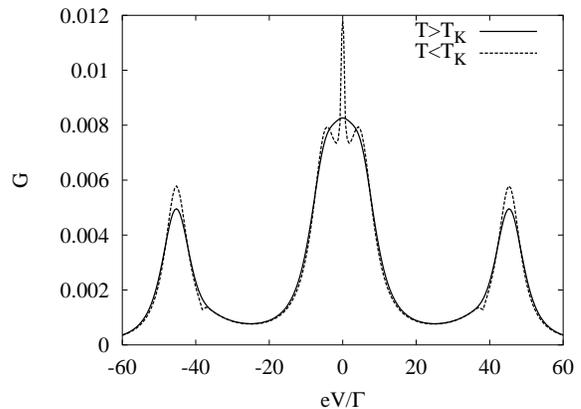}}
\caption{The differential conductance $G(V)$ versus
the applied bias $V$ for temperatures above and below
$T_{K}$ and a finite value of the Coulomb interaction 
$U=50 \Gamma$.}
\label{transport_fig2}
\end{figure}

\section{Small perturbations $V$ and $\Delta T$}

For infinitesimally small perturbations $V$ and $\Delta T$ 
one can expand the currents (\ref{charge_current},
\ref{heat_current}) up to linear terms
\begin{eqnarray}
I & = & -\frac{1}{T} L_{11} \; \nabla \mu 
+ L_{12} \; \nabla \left( \frac{1}{T} \right) ,
\\
I_{Q} & = & -\frac{1}{T} L_{21} \; \nabla \mu 
+ L_{22} \; \nabla \left( \frac{1}{T} \right) .
\end{eqnarray}
These coefficients $L_{ij}$ of the linear response theory can
be determined from the corresponding correlation functions in 
the Kubo formalism \cite{Mahan}. The zero bias conductance is 
then given by $G(0)\!=\!\frac{e^{2}}{T} \; L_{11}$, the Seebeck 
coefficient by $S \!=\! - \; \frac{1}{e \; T} \; \frac{L_{12}}{L_{11}}$ 
and the thermal conductivity by $\kappa \!=\! {1\over T^2} \left(
L_{22}-{L_{12}^2 \over L_{11}}\right)$. Coefficients $L_{ij}$ 
can also be derived directly from equations  (\ref{charge_current},
\ref{heat_current}) and they become functionals of the QD density 
of states via 
\begin{eqnarray}
L_{11}  =  \frac{T}{h} \int_{-\infty }^{\infty } 
d\omega  \; \Gamma(\omega) \; \left[ - \; \frac{\partial 
f(\omega)}{\partial \; \omega} \right] \; \rho_{d}
(\omega), \label{L_11}  \\
L_{12}  =  \frac{T^{2}}{h} 
\int_{-\infty }^{\infty } d\omega  \; \Gamma(\omega) 
\;  \left[ - \; \frac{\partial f(\omega)}{\partial \; T} 
\right] \; \rho_{d}(\omega) , \label{L_12} \\
L_{22}  =  \frac{T^{2}}{h} 
\int_{-\infty }^{\infty } d\omega  \; \Gamma(\omega) 
\; \omega \;  \left[ - \; \frac{\partial f(\omega)}
{\partial \; T} \right] \; \rho_{d}(\omega)  
\label{L_22}
\end{eqnarray}
where $\rho_{d}(\omega)$ and derivatives of the Fermi 
function are to be computed for $V\!=\!0$. From the 
Onsager relation we have $L_{21}=L_{12}$.

From the equation (\ref{L_11}) we conclude that at very low 
temperatures the zero bias conductance $G(0)$ is proportional 
to the density of states $\rho_{d}(\omega\!=\!0)$. Appearance 
of the Kondo automatically enhances the zero bias conductance 
as has been indeed observed experimentally \cite{Goldhaber-98}. 
Moreover, for $T \rightarrow 0$ it reaches the unitary limit 
value $2e^{2}/h$. With an increase of temperature there 
activated the scattering processes involving some 
higher energy sectors \cite{Pustilnik-04}. 

The linear response relations (\ref{L_11}-\ref{L_22}) are 
restricted to the external voltage $V$ and/or temperature 
difference $\Delta T$ being small (in terms of $\Gamma$). 
In the remaining part of this section we give a brief 
summary of our numerical computations while the next 
section will be devoted to arbitrary large perturbations.

\subsection{Numerical results}

Using the linear response relations (\ref{L_11}-\ref{L_22})
we explored numerically the transport properties for several 
values of $T$, $U$, $\varepsilon_{d}$ keeping a fixed number  
of charge on the QD $\langle n_{d} \rangle = 0.5$. We assumed 
the flat coupling $\Gamma_{\beta}(\omega)=\Gamma$ for energies 
$|\omega| \leq D$ and $\Gamma_{\beta}(\omega)=0$ elsewhere.

%
\begin{figure}
\centerline{\epsfxsize=7cm \epsffile{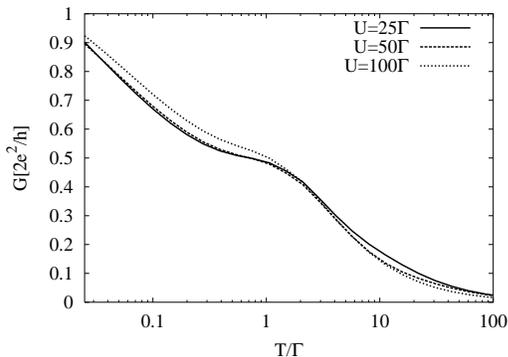}}
\caption{Temperature dependence of the zero bias differential 
conductance $G(V\!=\!0)$ obtained from the linear response 
calculations using the coefficient $L_{11}$ (\ref{L_11}).}
\label{cond_vs_T}
\end{figure}

Figure \ref{cond_vs_T} shows the temperature dependence of the 
zero bias conductance. We notice that $G(0)$ decreases monotonously 
versus temperature. At low temperatures $T \!\leq\! 0.1 \Gamma$, 
conductance decreases because of a gradual disappearance of the 
Kondo peak. For higher temperatures beyond the Kondo regime ($T 
\sim \Gamma$) there is a bit of a plateau and then again there 
starts an exponential decrease with respect to $T$. This behavior 
seems to be universal and the Coulomb interaction $U$ has merely 
a marginal effect on it.
 
%
\begin{figure}
\centerline{\epsfxsize=8cm \epsffile{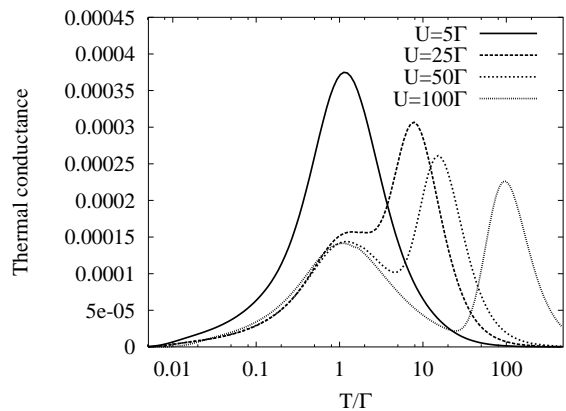}}
\caption{Temperature dependence of the thermal conductance 
$\kappa(T)$ obtained in the linear response theory.}
\label{transport_fig4}
\end{figure}

%
\begin{figure}
\centerline{\epsfxsize=7.5cm \epsffile{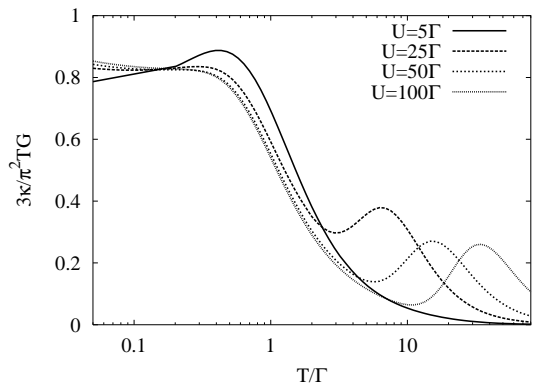}}
\caption{The Wiedemann Franz relation obtained in 
the linear response theory for the same values of $U$ 
as in figure \ref{transport_fig4}.}
\label{WF_law}
\end{figure}

On contrary, the effect of Coulomb interactions does show up in 
the temperature dependence of the thermal conductivity. In figure 
we can see \ref{transport_fig4} that at small temperatures 
$\kappa(T)$ exponentially increases versus $T$. For $T \sim \Gamma$ 
there appears one maximum (when the values of $U$ is small so that 
the Lorentzians presented in figure \ref{intro} overlap with one 
another) and then the other maximum occurs at $T \sim U$ (for 
sufficiently large $U$). The heat current is thus enhanced either 
when transport occurs through the low energy sector (the Lorentzian 
around $\varepsilon_{d}$) or it has an additional contribution by 
activating the high energy sectors (the Lorentzian around 
$\varepsilon_{d}+U$). We checked that the Kondo state has no 
influence on $\kappa(T)$. The mutual relation between $\kappa(T)$ 
and the $G(0)$ is summerized in figure \ref{transport_fig4}.

Figure \ref{transport_fig3} illustrates the Seebeck coefficient 
$S(T)$. At low temperatures $T<T_K$ the thermopower becomes negative 
due to the Kondo resonance slightly above the Fermi level (figure
\ref{intro}). A finite value of the Coulomb interaction $U$ leads 
to a partial flattening of the minimum in the Kondo regime. At 
higher temperatures $S(T)$ changes sign and considerably increases. 
For $T \sim U$ there occurs an other change of sign (negative 
$S(T)$) because of activating the high energy sector (the upper 
Lorentian in figure \ref{intro}). Similar behavior has been 
independently reported for the periodic Anderson model 
\cite{Pruschke-06}. At extremely large temperatures a magnitude 
of the thermopower asymptotically scales as $|S(T)| \propto T^{-1}$.

%
\begin{figure}
\centerline{\epsfxsize=7cm \epsffile{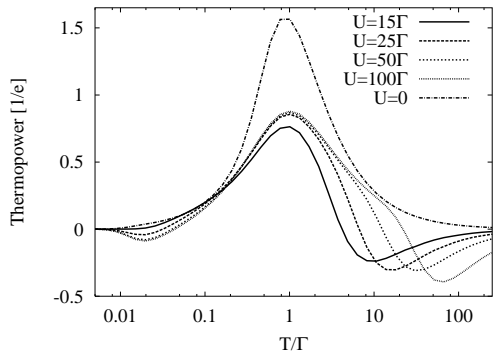}}
\caption{Temperature dependence of the thermopower $S(T)$ 
obtained in the linear response theory for several values 
of the Coulomb interaction: $U=50 \Gamma$ (dotted line), 
$U=25 \Gamma$ (dashed line) and $U=17.5 \Gamma$ (solid 
curve).}
\label{transport_fig3}
\end{figure}

\subsection{The Mott relation}

Within the linear response limit (i.e.\ for small perturbations 
$V$ and $\Delta T$) the following Mott formula \cite{Mott} 
\begin{eqnarray}
S^{M}(T) = -\; \frac{\pi^{2}}{3} \; \frac{k_{B}^{2}T}{e} \;
\frac{\partial \; \mbox{ln} G(0)}{\partial \mu} 
\label{Mott_formula}
\end{eqnarray}
is often used in experimental studies of the thermopower $S(T)$. 
Dependence of the zero bias conductance $G(V\!=\!0)$ on the chemical 
potential can be in practice masured by varrying the gate voltage 
$V_{G}$. Since the gate voltage shifts the energy levels of 
the QD one can assume that $- \; \partial \; \mbox{ln} G(0) / 
\partial \mu \simeq \partial \; \mbox{ln} G(0) / \partial V_{G}$. 

For the realistic multilevel QDs the Mott relation (\ref{Mott_formula}) 
implies for the thermopower to have a sawtooth shape as a function of 
$V_{G}$ \cite{Andreev01}. Similar property can be observed even 
in the single level QD as a consequence of the Coulomb blockade
\cite{Lunde-05}. The dashed lines in figure \ref{figure5} show 
the variation of the thermopower obtained from the Mott relation 
(\ref{Mott_formula}) for temperatures higher (the upper panel) 
and lower than the Kondo temperature $T_{K}$ (the bottom 
panel). In the Kondo regime we notice a clear discrepancy 
between the Mott formula and the results computed directly 
from (\ref{charge_current}) and (\ref{S}).

\begin{figure}
\centerline{\epsfxsize=7cm \epsffile{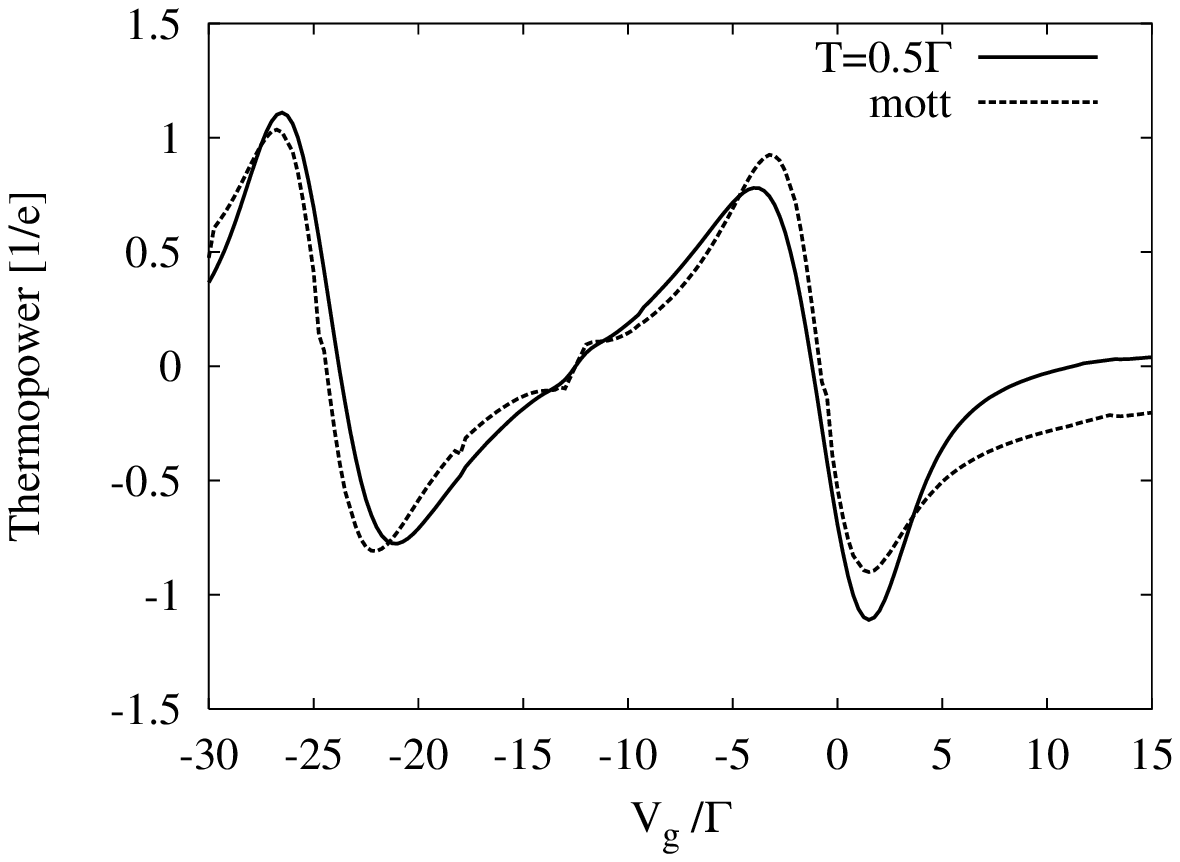}}
\centerline{\epsfxsize=7cm \epsffile{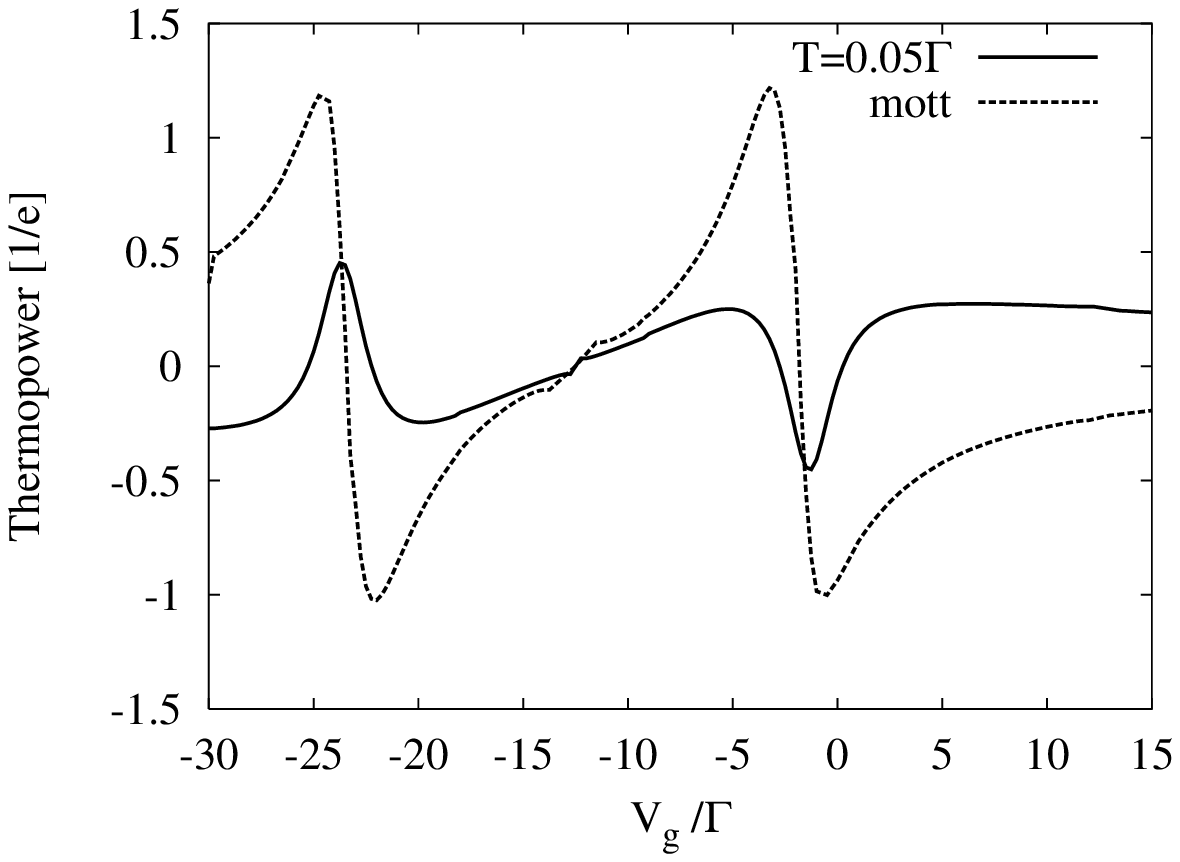}}
\caption{The thermopower $S(T)$ obtained directly from 
the linear response theory (solid lines) and using 
the Mott relation (\ref{Mott_formula}) (dashed lines).}
\label{figure5}
\end{figure}

In order to clarify a limitted range of applicability for 
the semi-classical Mott relation we briefly recollect some basic 
constraints used for deriviation of (\ref{Mott_formula}) as has 
been also independently stressed by other outhors \cite{Lunde-05}. 
The Landauer-type expression (\ref{charge_current}) can be expanded 
with respect to the small perturbations $V$ and $\Delta T$. If 
one neglects the temperature dependence of $\rho_{d}(\omega)$ 
(which obviously is not valid for $T \sim T_{K}$) then 
such expansion is carried out only in the Fermi distribution 
functions $f_{\beta}(\omega)$ 
\begin{eqnarray}
f_{\beta}(\omega) =  f(\omega) - \frac{\partial
f(\omega)}{\partial\;\omega} \left[ 
(\mu_{\beta}\!-\!\mu)  - \frac{\omega\! -\!\mu}{T} 
(T_{\beta}\!-\!T ) \right] .
\label{expansion}
\end{eqnarray}
Substituting $\mu_{L}\!=\!\mu\!+\!\frac{1}{2}eV$ and $\mu_{R}
\!=\!\mu\!-\!\frac{1}{2}eV$ together with (\ref{expansion}) into  
the equation (\ref{charge_current}) we search for such a bias $V$ 
which compensates the current induced by the temperature imbalance 
$\Delta T=T_L\!-\!T_R$. Using the definition (\ref{S}) we finally 
obtain \cite{Lunde-05}
\begin{equation}
S(T)= \frac{\int_{-\infty}^{\infty} 
d\omega (\omega-\mu) \Gamma(\omega) \rho_{d}(\omega)
\left[ - \frac{\partial f(\omega)}{\partial \omega}\right]} 
{eT \; \int_{-\infty}^{\infty} d\omega  \Gamma(\omega) 
\rho_{d}(\omega) \left[ - \frac{\partial 
f(\omega)}{\partial \omega}\right]}.
\label{ala_Lunde}
\end{equation}
For a quantitative determination of (\ref{ala_Lunde}) one can 
apply the Sommerfeld expansion $\int_{-\infty}^{\infty} d\omega 
f(\omega) Y(\omega) \simeq \int_{-\infty}^{\mu} d\omega 
Y(\omega)+\frac{\pi^{2}}{6}(k_{B}T)^2 Y^{'}(\mu)$ which 
is valid only at low temperatures. Within the lowest order 
estimation one finally gets
\begin{equation}
S(T) = -\; \frac{\pi^{2}}{3} \; \frac{k_B^{2}T}{e} \;\; 
\frac{\partial \; \mbox{ln} \rho_{d}(\omega\!=\!0)_{|_{T=0}}}
{\partial \mu} 
\end{equation}
because  $\lim_{T\rightarrow 0} \frac{\partial f(\omega)}
{\partial \omega}\simeq -\;\delta(\omega\!-\!\mu)$. 
According to (\ref{L_11}) the zero bias conductance  
$\lim_{T\rightarrow 0} G(V\!=\!0) = \mbox{const} \;
\times \; \rho_{d}(\omega\!=\!0)_{|_{T=0}}$
hence we finally arrive at the relation (\ref{Mott_formula})
\begin{eqnarray}
S(T) = -\; \frac{\pi^{2}}{3} \; \frac{k_B^{2}T}{e} \;\;
\frac{\partial \; \mbox{ln} G(V\!=\!0)_{|{_{T=0}}}}{\partial \mu} 
\label{modified_Mott}
\end{eqnarray}
It is now clear that deviation from the Mott relation shown 
in figure \ref{figure5} are due to the temperature dependence 
of the QD spectrum $\rho_{d}(\omega)$ when the Kondo peak 
gradually builds in.

\begin{figure}
\centerline{\epsfxsize=7cm \epsffile{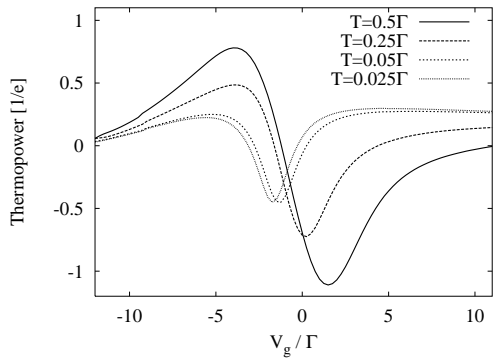}}
\caption{The thermopower $S(T)$ computed within the linear 
response theory in a range of the gate voltage $V_{G}$ where 
the Kondo effect has the most strong influence.}
\label{figure6}
\end{figure}

In figure \ref{figure6} we present the qualitative changes 
of the thermopower caused by the Kondo effect. For high 
temperatures $T\!>\!T_{K}$ we see a piece of the saw-tooth 
behavior \cite{Andreev01}. Upon lowering the temperatures 
to $T\!<\!T_{K}$ the Kondo resonance is formed in the 
density of states (see figure \ref{intro}) and through 
the relation (\ref{ala_Lunde}) this affects a sign of 
the thermopower. Such physical effects have been recently 
observed experimentally \cite{Scheibner-05}.

It is interesting to notice that the qualitative changes 
of the thermopower occur on both teeth of the {\em saw}.
We explain in the appendix that this is related to the 
particle or hole type of the Kondo resonance.

\section{Beyond the linear response}

%

We also studied the charge (\ref{charge_current}) and heat currents 
(\ref{heat_current}) going beyond the limit of small bias $V$ and 
temperature difference $\Delta T$. Figure \ref{transport_fig2} shows 
the differential conductance computed numericaly for $T_{L}\!=\!T_{R}$. 
When temperatures $T_{\beta}$ differ from one another the differential 
conductance has still qualitatively the similar behaviour with only the 
zero bias anomaly being gradually smeared for increasing $\Delta T$.

To account for nolinear effects of the other transport quantities 
we determied the thermopower and thermal conductivity directly from 
the definitions (\ref{S}-\ref{kappa}). Following \cite{Boese-01}, 
we solved numerically the integral equation
\begin{eqnarray}
I(V,\Delta T) = 0
\end{eqnarray}
for the fixed temperature difference $\Delta T$  determining the 
bias $V(\Delta T)$. The thermopower is then given by the ratio 
(\ref{S}). To compute the thermal conductance we determined 
the heat current $I_{Q}(V,\Delta T)$ and substituted it to 
(\ref{kappa}) using $V(\Delta T)$.

%
\begin{figure}
\centerline{\epsfxsize=7cm \epsffile{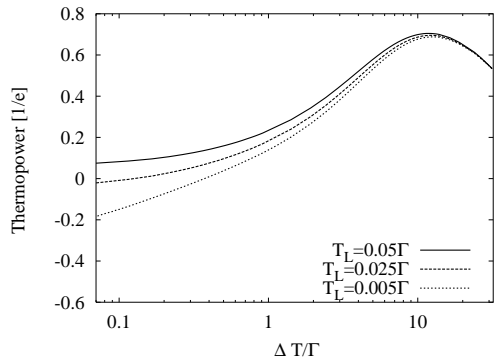}}
\caption{The thermopower $S(T)\equiv \frac{V}{\Delta T}_{|_{I=0}}$ 
determined beyond the linear response theory in the Kondo regime 
for $\varepsilon_{d}=-3 \Gamma$ and $U=50 \Gamma$.}
\label{figure7}
\end{figure}

In figure \ref{figure7} we show influence of the finite temperature 
difference $\Delta T$ on the thermopower $S(T)$ as function of the 
average temperature $T=(T_{L}+T_{R})/2$.  Our results prove that
the nonlinear effects play a considerable role. In general, by 
increasing the temperature difference $\Delta T$ we notice that 
the system acquires such properties which are characteristic for 
the high temperature behaviour shown previously in figure 
\ref{transport_fig3}.

%
\begin{figure}
\centerline{\epsfxsize=7cm \epsffile{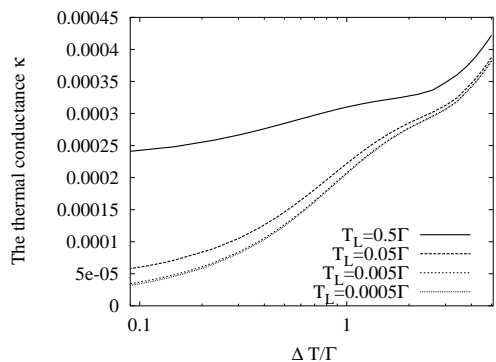}}
\caption{Variation of the thermal conductance $\kappa(T)$ as 
a function of the applied temperature difference  $\Delta T$
at low temperature region (upper panel) and at higher
temperatures (the bottom panel).}
\label{figure9}
\end{figure}

Thermal conductance $\kappa$ is a physical quantity which
is rather weakly sensitive to any particular structure of
the low energy excitations. In the linear response limit it
is known to vanish for $T\rightarrow 0$ as well as for the 
opposite limit $T\rightarrow 0$ \cite{Boese-01,Krawiec-02}.
The maximum value of $\kappa$ occurs usually at temperatures 
$T \sim \Gamma/k_{B}$ (see figure \ref{transport_fig4}). This 
situation changes radically if the applied temperature difference
$\Delta T$ is large. Obviously, $\kappa(T)$ gradually increases 
as a function of $\Delta T$ (see the figure \ref{figure9}) but
this dependence is irregular. We further illustrate it in
figure \ref{figure10} where we plot the Wiedemann-Franz ratio
versus the average temperature $T$ for several fixed values
of $T_{L}$. 

\begin{figure}
\centerline{\epsfxsize=7cm \epsffile{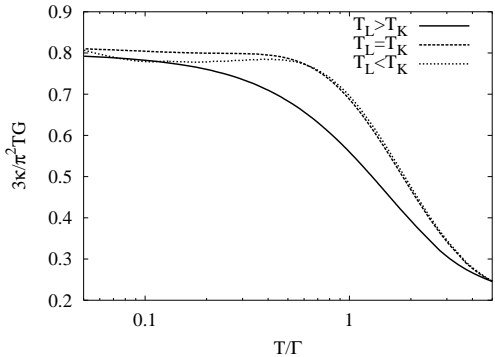}}
\caption{The Wiedemann-Franz relation beyond the linear response
limit determined for several values of $T_{L}$ with the Coulomb 
energy $U=25 \Gamma$. On abscissa we put the everage temperature 
defined as $T=T_L+\frac{\Delta T}{2}$.}
\label{figure10}
\end{figure}

\section{Summary}

We studied the influence of correlations on various properties 
of the charge and heat transport through the quantum dot connected 
to the normal leads. In the regime of linear response theory (for 
the small applied bias $V$ or temperature difference $\Delta T$) 
we find the qualitative changes caused strictly by appearance of 
the Kondo resonance at low energies. In particular, it leads to 
the zero bias anomaly and to the change of sign of the thermopower.
Moreover, the Coulomb interactions have additional effects
by activation of the high energy channels in the QD spectrum
for the transport of charge and heat. Presence of such high 
energy states manifests by the nonmonotonous variation of 
the thermopower (figure \ref{transport_fig3}) and thermal 
conductance (figure \ref{transport_fig4}). 

Appearance of the Kondo effect has a further strong influence
on deviation from the semiclassical Mott-relation. The Mott
formula (\ref{Mott_formula}) fails for $T \sim T_{K}$ because 
of the temperature dependence of  the spectral function. 

The Wiedemann-Franz law seems to be rather well preserved 
unless temperatures are small, for incresing temperatures 
we observe a systematic departure from the well known 
Wiedemann-Franz ratio
The nonlinear effects studied within the Landauer formalism
show an additional effect on the transport properties for
practically all transport quantities. This issue has been 
previously pointed also by some other authors \cite{Boese-01}.

\acknowledgements 

We kindly acknowledge fruitful discussions with M.\ Krawiec. This 
work is partially supported by the KBN grant No.\ 2P03B06225.

\appendix*
\section{}

In this appendix we briefly explain under what conditions 
there can arise the Kondo effect in the system described
by Hamiltonian (\ref{model}). For this purpose let us first 
explore the particle hole transformation
\begin{eqnarray}
d_{\sigma} \rightarrow \tilde{d}_{\sigma}^{+}
&\hspace{1cm} &
c_{{\bf k}\beta \sigma} \rightarrow 
\tilde{c}_{{\bf k}\beta \sigma}^{+} 
\nonumber \\
d_{\sigma}^{+} \rightarrow \tilde{d}_{\sigma}
&&
c_{{\bf k}\beta \sigma}^{+} \rightarrow 
\tilde{c}_{{\bf k}\beta \sigma}
\nonumber
\end{eqnarray}
Substituting these operators into the Hamiltonian (\ref{model}) 
we obtain
\begin{eqnarray}
H & = & 
\sum_{{\bf k},\beta,\sigma } \tilde{\xi}_{{\bf k}\beta}
\tilde{c}^{+}_{{\bf k} \beta \sigma} \tilde{c}_{{\bf k} \beta \sigma}+
\sum_{\sigma} \tilde{\epsilon}_{d} \tilde{d}^{+}_{\sigma}\tilde{d}_{\sigma} 
+  \tilde{U} \tilde{n}_{d\uparrow} \tilde{n}_{d\downarrow} \nonumber \\
& + & \sum_{{\bf k},\beta, \sigma} \left( 
\tilde{V}_{{\bf k} \beta} \tilde{c}_{{\bf k} \beta \sigma}
\tilde{d}_{\sigma}^{+} + \tilde{V}_{{\bf k} \beta}^{*} 
\tilde{c}^{+}_{{\bf k} \beta \sigma} \tilde{d}_{\sigma} \right) 
+ const 
\nonumber
\end{eqnarray}
where 
\begin{eqnarray}
\tilde{\xi}_{{\bf k}\beta} & = & - \; \xi_{{\bf k}\beta} 
\label{A1}\\
\tilde{\epsilon}_{d} & = & - \; \epsilon_{d} - U 
\label{A2}\\
\tilde{U} & = & U \label{A3}\\
\tilde{V}_{{\bf k} \beta} & = & -\; V_{{\bf k} \beta}^{*}
\label{A4}\end{eqnarray}
and $const=U\!+\!2\varepsilon_{d}\!+\!\sum_{{\bf k},\beta,\sigma}
\xi_{{\bf k}\beta\sigma}$. We thus notice that the Hamiltonian 
(\ref{model}) preserves its structure under the particle-hole 
transformation and simultaneously the model parameters are scaled
as given in (\ref{A1}-\ref{A4}). This property will be useful
for us when determining conditions necessary for the Kondo effect
to appear.

The standard way to specify when electron of the QD is
perfectly screened by the mobile electrons is obtained using
the Schrieffer-Wolf transformation. This perturbative method 
eliminates the hybridization terms $V_{{\bf k}\beta}$ to linear 
terms. In the present case one derives the following
effective superexchange coupling \cite{Bruus_book}
\begin{eqnarray}
J = \frac{U \; |V_{k_{F},\beta}|^{2}}{(\varepsilon_{d}+U)
(-\varepsilon_{d})} .
\label{superexchange_coupling}
\end{eqnarray}
In the equilibrium case (i.e.\ for $V\!=\!0$) this formula
(\ref{superexchange_coupling}) means that antiferromagnetic 
interactions arise when the energy level $\varepsilon_{d}$ 
of the QD is located slightly below the chemical potential 
(we assumed here $\mu\!=\!0$). This situation is schematically
illustrated in figure 1.

The other possibility for the Kondo state to appear corresponds
to holes rather than particles. In a simple way one can verify
that under the particle hole transformation the superexchange  
coupling transforms to
\begin{eqnarray}
J = \frac{\tilde{U} \; |\tilde{V}_{k_{F},\beta}|^{2}}
{(-\tilde{\varepsilon}_{d})(\tilde{\varepsilon}_{d}+\tilde{U})} .
\label{superexchange_forholes}
\end{eqnarray}
This result could be expected on more general grounds due
to the fact that the XY model is invariant on the particle
hole transformation. We hence conclude that the Kondo effect 
should be present for the small negative value of $\tilde{
\varepsilon}_{d}$, in other words for $\varepsilon_{d}$ 
located slightly above $U$. We would like to emphasize 
that this is not an artifact depending on the 
approximations used in this work.

\end{document}